\pdfoutput=1

\documentclass[conference]{IEEEtran}
\usepackage{graphicx}
\usepackage{amssymb}
\usepackage{bm}
\usepackage{amsmath}
\usepackage{pgfplots}
\usepackage{balance}
\usepackage{booktabs} 
\pgfplotsset{compat=newest}
\pgfplotsset{plot coordinates/math parser=false}
\newlength\figureheight
\newlength\figurewidth

\usepackage{color,soul}
\IEEEoverridecommandlockouts

\ifCLASSOPTIONcompsoc
    \usepackage[caption=false, font=normalsize, labelfont=sf, textfont=sf]{subfig}
\else
\usepackage[caption=false, font=footnotesize]{subfig}
\fi
  
\newtheorem{remark}{Remark}
\newcommand{\trm}[1]{\textrm{#1}}
\newcommand{\Figref}[1]{Figure~\ref{#1}}

\newcommand{\inv}{^{-1}}

\begin{document}

\title{Utility-based Downlink Pilot Assignment in Cell-Free Massive MIMO}
\author{\IEEEauthorblockN{Giovanni Interdonato$^{*\dagger}$, P{\aa}l Frenger$^*$, Erik G. Larsson$^\dagger$}
\IEEEauthorblockA{$^*$Ericsson Research, 581 12 Link\"oping, Sweden\\
$^\dagger$Department of Electrical Engineering (ISY), Link\"oping University, 581 83 Link\"oping, Sweden\\
\{giovanni.interdonato, pal.frenger\}@ericsson.com, erik.g.larsson@liu.se
\thanks{This paper was supported by the European Union's Horizon 2020 research
and innovation programme under grant agreement No 641985 (5Gwireless).}}}
\maketitle

\begin{abstract}
We propose a strategy for orthogonal downlink pilot assignment in cell-free massive MIMO (multiple-input multiple-output) that exploits knowledge of the channel state information, the channel hardening degree at each user, and the mobility conditions for the users. These elements, properly combined together, are used to define a \textit{user pilot utility} metric, which measures the user's real need of a downlink pilot for efficient data decoding. The proposed strategy consists in assigning orthogonal downlink pilots only to the users having a pilot utility metric exceeding a predetermined threshold. Instead, users that are not assigned with an orthogonal downlink pilot decode the data by using the statistical channel state information. The utility-based approach guarantees higher downlink net sum throughput, better support both for high-speed users and shorter coherent intervals than prior art approaches.  
\end{abstract}

\section{Introduction}
Cell-free massive MIMO~\cite{ngo2017cellfree} comes out from the combination of massive MIMO~\cite{redbook}, distributed architecture, and cell-less network topology. In such a system, a very large number of base station antennas, named as access points (APs) herein, are geographically distributed and connected to one (or more) central process unit (CPU) through a fronthaul network. All the APs jointly and coherently serve a smaller number of users in the same time-frequency resources. 

The use of coherent cooperation as key enabling to mitigate the inter-cell interference has been fully exploited in literature, and it results in concepts known as network MIMO~\cite{huh2012network}, coordinated multipoint with joint processing (CoMP-JP)~\cite{irmer2011comp}, and multi-cell MIMO cooperative networks~\cite{gesbert2010mumimocooperative}. Cell-free massive MIMO can be considered the scalable way to implement network MIMO, since channel state information (CSI) acquisition and channel estimation can be performed locally at each AP by exploiting the \textit{channel reciprocity} property of time-division duplex (TDD) systems~\cite{bjornson2010cooperative}.   

Compared to conventional (cellular centralized) massive MIMO, cell-free massive MIMO guarantees an increased macro-diversity gain, since each UE receives the same signal from very different paths. Moreover, the co-processing over the APs alleviates the interference and yields higher spectral efficiency to cell-edge users~\cite{ngo2017cellfree}. Hence, each user experiences that there are no cell boundaries since it is surrounded by many serving APs (hence the term cell-free). However, these benefits come at the price of increased fronthaul network requirements. Furthermore, cell-free massive MIMO is characterized by a low degree of \textit{channel hardening}~\cite{interdonato2016dlpilot,chen2017channelhardening}.

The channel hardening property is crucial in conventional massive MIMO to eliminate the effects of small-scale fading. The channel variations averaged out when the same signal is transmitted by a massive number of co-located antennas over multiple stochastic channels. This is a direct consequence of the \textit{law of the large numbers}. Thanks to the channel hardening, the user sees a channel that behaves almost as constant, and it can reliably decode data by using only long-term statistical CSI. Hence, downlink training that facilitates the users to acquire instantaneous CSI is not needed in massive MIMO~\cite{ngo2017nodownlinkpilots}. 

In contrast, in cell-free massive MIMO the channel hardening is, in general, less pronounced due to the distributed architecture. Basically, only a subset of APs, i.e., the closest APs, effectively serve a given user because the other contributes are attenuated by larger path-loss. Hence, the number of APs actually involved in the transmission is smaller and the law of the large numbers is not always applicable. Consequently, users with low degree of channel hardening need to estimate the CSI in a short-term basis from downlink pilots~\cite{interdonato2016dlpilot}. High-speed users also need downlink pilots to acquire instantaneous CSI, regardless of the channel hardening, because of the fast variations of their channel. However, downlink pilot transmission requires further radio resources which are subtracted from the data. Hence, a resource-saving approach while guaranteeing better system performance is needed when assigning the downlink pilots to the users.

\textbf{Contributions:} We propose an utility-based strategy for orthogonal downlink pilot assignment in cell-free massive MIMO, and provide different variants of pilot utility metrics as function of user speed, channel hardening degree, user priority, and rate estimates. We give an analytic definition of channel hardening degree for cell-free massive MIMO systems with i.i.d. Rayleigh fading channel. A performance comparison between the proposed scheme and the prior art is provided. This study focus to the case of mutually orthogonal pilots. The case with pilot reuse will be investigated in future work.  

\section{Pilot assignment in Massive MIMO}


Acquiring high quality CSI to facilitate phase-coherent processing at multiple antennas is a crucial activity in massive MIMO.  
By operating in TDD mode, massive MIMO benefits from the channel reciprocity property, according to which the channel responses are the same in both uplink (UL) and downlink (DL). The BS can estimate the channels from a-priori known signals, called \textit{pilots}, transmitted by the UEs in the UL (UL training). Thanks to the channel reciprocity, these estimates are then valid for both the UL and the DL. Similarly, in the DL, the UEs may need to estimate the channel from DL pilots sent by the BS (DL training). 

\subsection{TDD protocol and Resource Allocation}
\begin{figure}[!t]
    \centering
    \subfloat[TDD frame including the DL training]{\includegraphics[width=.95\linewidth]{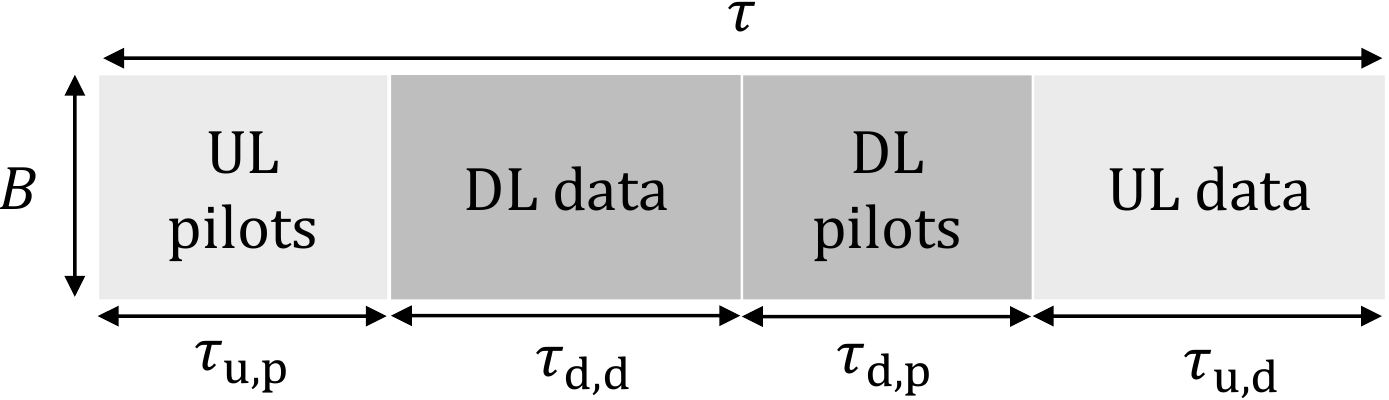}
    \label{fig:dlpilots}}\\     
    \subfloat[TDD frame with no DL training]{\includegraphics[width=.95\linewidth]{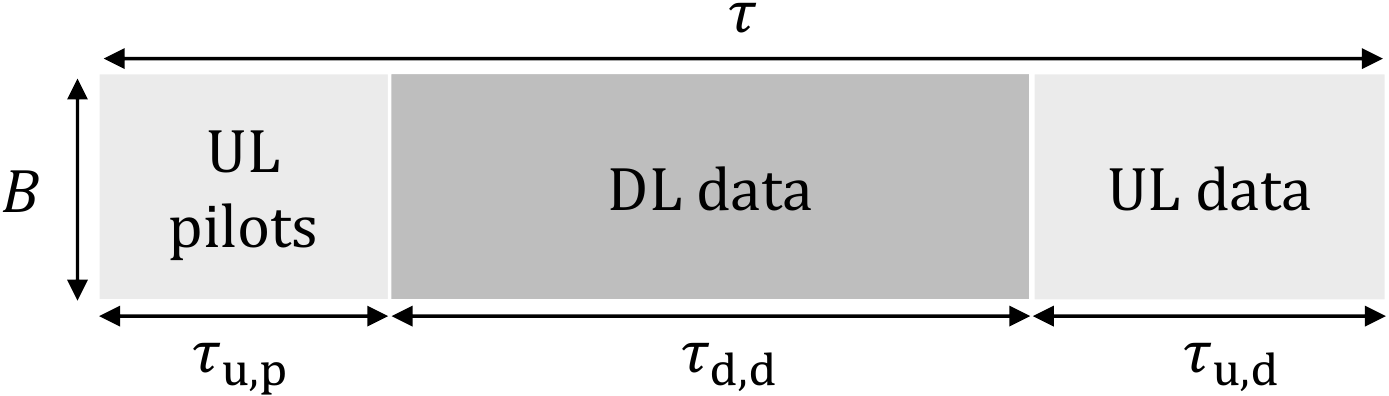}
    \label{fig:ulpilots}}        
    \caption{The TDD frame consists of $\tau = T_\trm{c}B$ symbols, where $B$ and $T_c$ are the coherence bandwidth and the coherence time, respectively.}
    \label{fig:TDDframe}
\end{figure}

The TDD frame is generally designed to be smaller or equal to the channel coherence interval of all UEs. Here, we assume that a TDD frame is equal to the channel coherence interval, and that they are interchangeable terms. A TDD frame may consist of either three or four phases whether the DL training is performed or not, respectively. The TDD frame including the DL training phase consists of: $(i)$ UL training (UL pilots); $(ii)$ DL payload data transmission (DL data); $(iii)$ DL training (DL pilots); and $(iv)$ UL payload data transmission (UL data), as shown in \Figref{fig:dlpilots}. The TDD frame structure for the case with no DL training is shown in \Figref{fig:ulpilots}. 
The TDD frame is $\tau = T_\trm{c}B$ symbols long, where $B$ indicates the coherence bandwidth and $T_\trm{c}$ is the coherence time. Let $\tau_{\trm{p}}$ denote the total number of symbols per TDD frame spent on transmission of pilots, $\tau_{\trm{d,d}}$ and $\tau_{\trm{u,d}}$ denote the number of symbols per TDD frame spent on transmission of DL and UL payload data, respectively. Hence, the length of the TDD frame is given by $\tau = \tau_{\trm{p}} + \tau_{\trm{d,d}} + \tau_{\trm{u,d}}$.

\begin{figure}[!t]
    \centering
    \includegraphics[width=.48\textwidth]{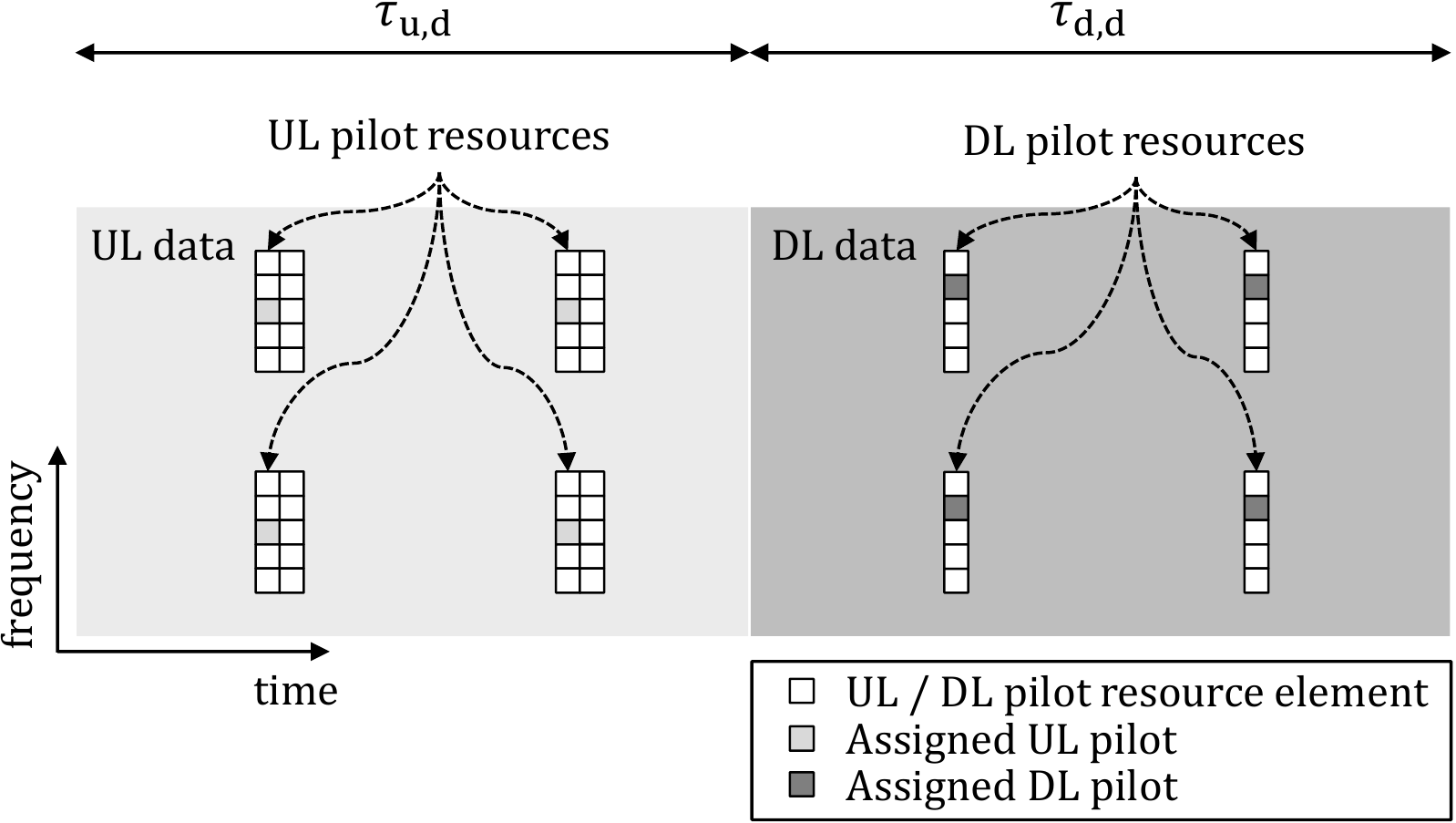}
    \caption{The data and pilot transmissions are typically interleaved in the time-frequency grid. In this example, 10 orthogonal UL pilots consumes 40 REs in the UL data (e.g., 1 pilot consists of 4 REs). The DL data contains 5 pilots consuming 20 REs.}\label{fig:regrid}
\end{figure}

In many practical systems, the pilot and data transmissions are interleaved in the time-frequency domain, as depicted schematically in \Figref{fig:regrid}. Assuming a time-frequency grid of resource elements as in the 3GPP standards LTE and the upcoming New Radio (NR), a UL pilot (DL pilot) may consist of several resource elements (REs). Data as well as pilots in UL (DL) are transmitted during the UL data (DL data) phase.
In the example in \Figref{fig:regrid}, there are 10 UL pilots and 5 DL pilots, each consisting of 4 REs. If the UL and DL pilots have to be orthogonal in order not to generate interference, the pilot REs cannot be used for data transmission by any other user. In addition, the assignment of UL and DL pilot resources is semi-static. Consequently, even when there is only one single UE active in a UL or DL data the REs reserved for the UL or DL pilots cannot be allocated for different purposes. Adapting the number of REs reserved for the pilots to the dynamic changes of the network requires a significant amount of control signaling which both consumes radio resources and introduces delay. Therefore, minimizing the pilot overhead without affecting the performance represents a relevant aspect when designing the transmission protocol.

\subsection{Pilot Contamination and Orthogonal Pilot Transmission} 

Massive MIMO performance is profoundly affected by a basic phenomenon named as \textit{pilot contamination}~\cite{jose2011pilotcontamination}. Pilot contamination arises when two or more UEs send the same, or in general non-orthogonal, pilot sequences. More specifically, sending non-orthogonal pilots causes mutual interference which deteriorates the quality of the corresponding channel estimates. In addition, pilot contamination does not vanish as the number of BS antennas grows large, thus it is an impairment that remains asymptotically. Ideally, every UE in a massive MIMO system should be assigned an orthogonal UL pilot sequence in order not to create interference and contaminate the UL channel estimates. However, the maximum number of orthogonal pilot sequences that can be assigned to the UEs is upper-bounded by the number of symbols in the channel coherence interval. Since this number is finite, and depending on propagation environments, the assignment of mutually orthogonal UL pilots might be physically impossible.

The radio resources become even more limited if DL pilots need to be allocated. As mentioned earlier, in massive MIMO the DL pilots are not needed because they do not introduce relevant gain in term of system capacity, by virtue of the combined effect of channel hardening and channel reciprocity. However, this holds under the assumption of low/moderate-speed user. High-speed UEs suffer from inaccurate and outdated CSI (CSI aging) as their channel conditions change very quickly. Therefore, they need to be trained in the DL to know the instantaneous CSI, regardless of the channel hardening. Conversely, DL training is also in general more beneficial in cell-free massive MIMO due to the lower degree of channel hardening~\cite{interdonato2016dlpilot}. 

The DL training scheme introduced in~\cite{ngo2013dlpilots} for conventional massive MIMO, and repurposed in~\cite{interdonato2016dlpilot} for cell-free massive MIMO, consists in assigning one orthogonal DL pilot per UE, and using conjugate beamforming to send the pilot, instead of simply broadcasting. This allows to sensibly reduce both the mutual pilot interference and the pilot overhead, which now scales as the number of the UEs instead of the number of antennas. However, if the number of active UEs exceeds the number of symbols that can be afforded for DL pilots, mutually orthogonal DL pilots cannot be assigned, and pilot reuse is necessary. 

In this paper, we assume that only mutually orthogonal pilots can be assigned. If the number of the active UEs is larger than the number of symbols per frame reserved for pilots, then the exceeding users will not be served in the current frame but scheduled in the next frame. Adopting orthogonal pilots with no pilot reuse eliminates the pilot contamination and:
\begin{itemize}
\item makes the interference from DL signals towards different users possible to identify and report. If a UE experiences a large amount of DL interference, it can try to correlate the received signal with the different candidate pilot sequences and identify the interference source. The interfered UE may then report the interfering pilot index to the serving BS (or APs) which can eliminate the interference, e.g. by using more robust link adaptation when both interfered and interfering UEs are active, or by separating the UEs in time/frequency;
\item enables a simpler overall implementation of the system, since no special measures need to be taken to handle the pilot contamination. The handling of the potential error cases associated with pilot reuse or non-orthogonal pilots increases the complexity at the BS/AP;
\item reduces the overhead due to the pilot reconfiguration process. When pilots are reused, the BS (or the APs) may repeatedly re-assign the UL/DL pilots to reduce the pilot contamination between UEs spatially correlated (e.g., assigning orthogonal pilots to neighbor UEs and reusing the same pilot for distant UEs). With orthogonal pilots, each UE can keep the same pilot during its lifetime in the network. Hence, orthogonal pilots are \textit{UE specific}.
\end{itemize}

\section{System Model} \label{sec:sysmodel}

\begin{figure}[!t]
    \centering
    \includegraphics[width=1\linewidth]{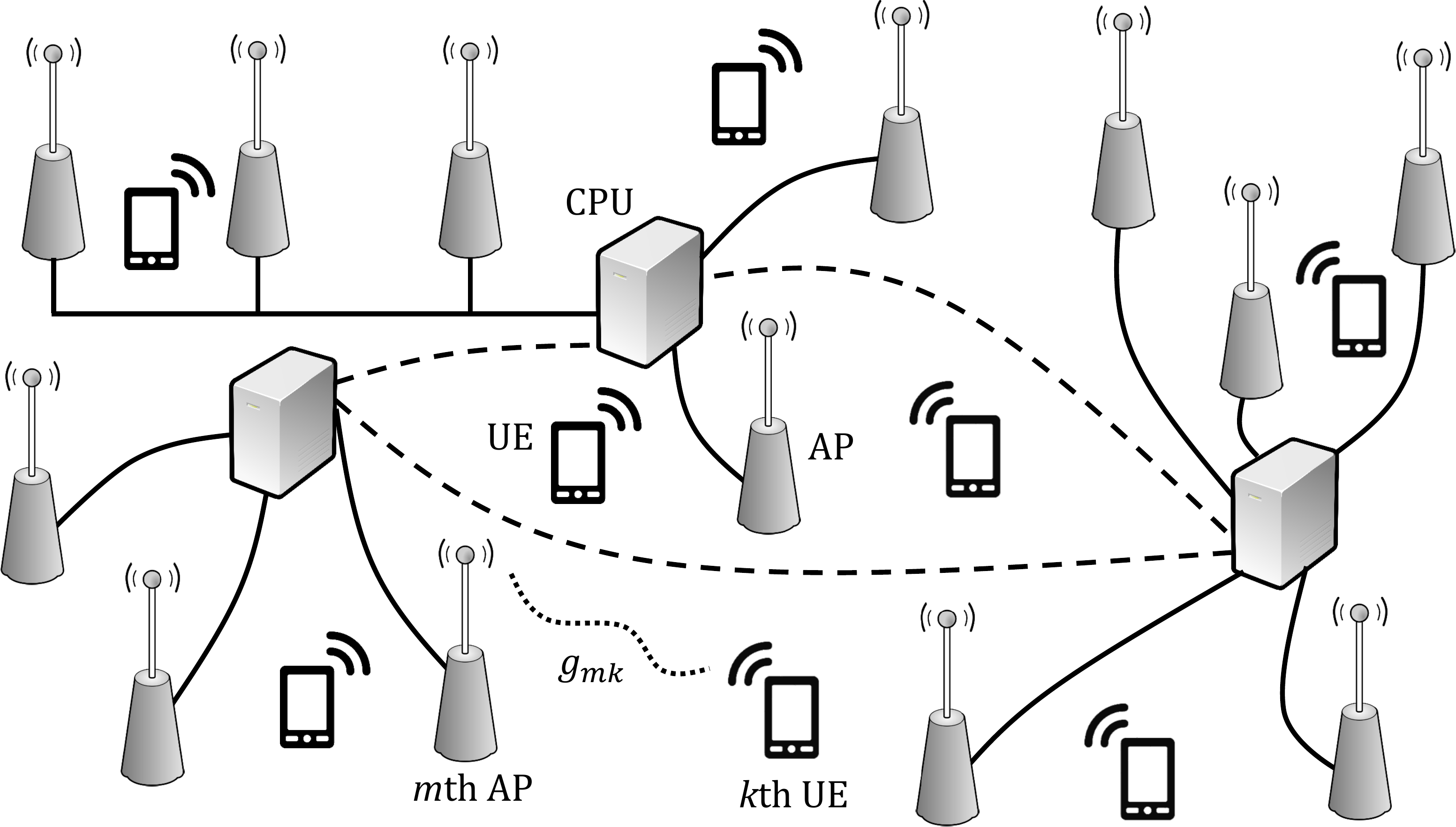}
    \caption{In cell-free massive MIMO, a large number of APs serve simultaneously a smaller number of UEs. The CPUs handle the APs cooperation.}\label{fig:network}
\end{figure}

We consider a TDD cell-free massive MIMO system with $M$ single-antenna APs and $K$ single-antenna UEs, $M>K$. We assume both the APs and the UEs are randomly spread out in a large area without boundaries. The APs are connected to one or more CPUs through a fronthaul network with different network topologies, as shown in \Figref{fig:network}. 
Each AP locally acquires CSI and estimates the channel from mutually orthogonal UL pilots sent by the UEs. These estimates are used in the DL for precoding, assuming perfect channel reciprocity calibration. We focus on \textit{conjugate beamforming}, also known as \textit{maximum-ratio transmission}, as DL precoding technique. Although it does not represent the optimal precoder, performing such a linear processing offers low operational complexity with inexpensive hardware components. In addition, conjugate beamforming does not require CSI sharing among APs and CPU, with benefits on the fronthaul network load. Therefore, the CPU is responsible to collect and elaborate only payload data, and to implement advanced techniques for DL power control and DL/UL pilot assignment.

The TDD frame length is equal to the coherence interval. Hence, the channel is assumed to be static within a frame and variable independently for each frame. Let $g_{mk}$ denote the channel gain between the $m$th AP and the $k$th user (and vice versa) defined as 
\begin{equation}
\label{eq:channelmodel}
g_{mk} = \sqrt{\beta_{mk}}h_{mk} \sim \mathcal{CN}(0,\beta_{mk}),
\end{equation}
where $h_{mk}$ represents the small-scale fading, and $\beta_{mk}$ is the large-scale fading. We assume that $\{h_{mk}\}$ coefficients follow a Rayleigh distribution, i.e, $h_{mk} \sim \mathcal{CN}(0,1)$ and i.i.d. RVs, for $m = 1,\ldots,M$, $k = 1, \ldots, K$. The large-scale fading addresses path-loss and shadow-fading. We assume the large-scale fading coefficients are estimated a-priori and known whenever required.

The AP estimates the UL channel by correlating the received UL pilot signal with the corresponding known pilot sequence and performing MMSE. The resulting UL channel estimated from the orthonormal UL pilot sequence $\bm{\varphi}_k\in\mathbb{C}^{\tau_{\textrm{u,p}} \times 1}$, $\Vert\bm{\varphi}_k\Vert^2=1$, is Gaussian RV distributed as $\hat{g}_{mk}~\sim~\mathcal{CN}(0,\gamma_{mk})$ where 
\begin{equation} \label{eq:gamma}
\gamma_{mk} = \frac{\tau_{\textrm{u,p}}\rho_{\textrm{u,p}}\beta^2_{mk}}{\tau_{\textrm{u,p}}\rho_{\textrm{u,p}}\beta_{mk}+1},
\end{equation}
and $\rho_{\textrm{u,p}}$ represents the normalized transmit signal-to-noise ratio (SNR) associated to the UL pilot symbol.
\begin{remark}
The transmission of orthogonal UL pilots requires $\tau_{\textrm{u,p}} = K$. Each AP is able to resolve $K$ spatial dimensions since the channel estimates are orthogonal each other. The estimates are not affected by the pilot contamination.
\end{remark}
\begin{remark}
$\gamma_{mk}$ gives a measure of the estimation quality. In fact, $\gamma_{mk} \leq \beta_{mk}$, with equality if the estimation is perfect.
\end{remark}

In the DL, each AP serves all the UEs implementing power control, and using the UL estimates to perform conjugate beamforming. The data signal sent by the $m$th AP is
\begin{equation}
\label{eq:DLsignal}
x_m = \sqrt{\rho_\textrm{d}}\sum^K_{k=1} \sqrt{\eta_{mk}} \ \hat{g}^*_{mk} q_k,
\end{equation}
where $q_k$ is the unit-power data symbol, i.e., $\mathbb{E}\{\vert q_k \vert^2\}=1$, for the $k$th UE, and $\rho_\textrm{d}$
is the normalized transmit SNR related to the data symbol. The term $\hat{g}^*_{mk}$ represents the \textit{precoding factor}, and $\{\eta_{mk}\}$, $m=1,...,M$, $k=1,...,K$, are the power control coefficients satisfying the per-AP power constraint
\begin{equation}
\label{eq:pwConstraint}
\mathbb{E}\{|x_m|^2\}\leq\rho_\textrm{d} \implies \sum \limits_{k=1}^K \eta_{mk} \gamma_{mk} \leq 1, \ \text{for all}\ m.
\end{equation}
The effective DL channel gain seen by the $k$th UE is given by
\begin{align}\label{eq:akk}
a_{kk'} \triangleq \sum^M_{m=1} \sqrt{\eta_{mk'}} {g}_{mk}\hat{g}^*_{mk'}, ~ k'=1,...,K.
\end{align}
The $k$th UE can detect the data symbol $q_k$ only if it has a sufficient knowledge of $a_{kk}$.

\section{Achievable Downlink Rate}

In the DL, the UEs can either rely on statistical CSI or estimate the instantaneous CSI to perform data decoding. The DL channel can be estimated, as in the UL, from DL pilot beamformed by each AP, as described in \cite{interdonato2016dlpilot}.

\subsection{Fading Channel with Statistical CSI at the Receiver}
If each UE has knowledge of the channel statistics but not of the channel realizations, a lower bound on the DL ergodic capacity can be calculated as described in~\cite[Sec. 2.3.4]{redbook}, assuming that Gaussian codebooks are used.

An achievable DL rate of the transmission from the APs to the $k$th UE in the cell-free massive MIMO system with conjugate beamforming, mutually orthogonal UL pilots, for any finite $M$ and $K$, is given by~\cite{ngo2017cellfree} and equal to
\begin{equation} \label{eq:RsCSI}
R_k^{\textrm{sCSI}} = \log_2 \left( 1 + \frac{\rho_\textrm{d}\left(\sum\limits^M_{m=1} \sqrt{\eta_{mk}}\gamma_{mk}\right)^2}{\rho_{\textrm{d}} \varsigma_{kk} + \rho_{\textrm{d}} \sum\limits^K_{k' \neq k} \varsigma_{kk'} + 1}\right),
\end{equation}
where $\varsigma_{kk'} \triangleq \sum_{m=1}^M \eta_{mk'} \beta_{mk} \gamma_{mk'}, k'=1,\ldots ,K$. The term $\rho_{\textrm{d}} \varsigma_{kk}$ represents the so-called \textit{beamforming uncertainty gain}. It comes from the UEs' lack of the instantaneous CSI. Hence, this gain gives an indirect measure of the channel hardening: the more the channel hardens for UE $k$, the smaller $\varsigma_{kk}$ is.  
\subsection{Fading Channel with Side Information}
If each UE has imperfect knowledge of the instantaneous CSI, a lower bound on the DL ergodic capacity can be calculated by invoking the capacity-bounding technique for channel with \textit{side information} (i.e., the channel estimates) as in \cite{medard2000capacity} and \cite[Sec. 2.3.5]{redbook}. A valid approximation for such a DL ergodic capacity lower bound is given in \cite{interdonato2016dlpilot} by observing that the effective DL channel gains $\{a_{kk'}\},~k = 1,\ldots,K$, can be approximated as Gaussian RVs. Hence, an achievable DL rate of the transmission from the APs to the $k$th UE in the cell-free massive MIMO system with conjugate beamforming, mutually orthogonal UL and DL pilots, for any finite $M$ and $K$, is given by~\cite{interdonato2016dlpilot} as follows   
\begin{equation} \label{eq:RiCSI}
R_k^{\textrm{iCSI}}\!\approx\!\mathbb{E}\left\{\!\log_2\!\left(1\!+\!\frac{\rho_\textrm{d}|\hat{a}_{kk}|^2}{\rho_\textrm{d}\frac{\varsigma_{kk}}{\tau_\textrm{d,p}\rho_\textrm{d,p}\varsigma_{kk}\!+\!1}\!+\!\rho_{\textrm{d}}\!\sum\limits^K_{k' \neq k} \varsigma_{kk'}\!+\!1}\!\right)\!\right\},
\end{equation}
where $\rho_{\textrm{d,p}}$ represents the SNR associated to the DL pilot symbol, $\hat{a}_{kk}$ is the estimate of the effective DL channel gain $a_{kk}$, given by 
\begin{equation}
\label{eq:a_est}
\hat{a}_{kk}\!=\!\frac{\tau_{\textrm{d,p}}\rho_{\textrm{d,p}}\varsigma_{kk} a_{kk}\!+\!\sqrt{\tau_{\textrm{d,p}}\rho_{\textrm{d,p}}}\varsigma_{kk}w_{\textrm{dp,}k}\!+\!\sum\limits_{m=1}^M\!\sqrt{\eta_{mk}}\!\gamma_{mk}}{\tau_{\textrm{d,p}}\rho_{\textrm{d,p}}\varsigma_{kk}\!+\!1}.
\end{equation}
The expectation in~\eqref{eq:RiCSI} is taken with respect to the channel estimate $\hat{a}_{kk}$. In~\eqref{eq:a_est}, the second term in the numerator is Gaussian RV as $w_{\textrm{dp},k} \sim \mathcal{CN}(0,1)$ is AWGN at the $k$th UE during the DL pilot signaling.

\section{Utility-based DL Pilot Assignment}
\subsection{Proposed Scheme}
\begin{figure}[!t]
    \centering
    \includegraphics[width=1\linewidth]{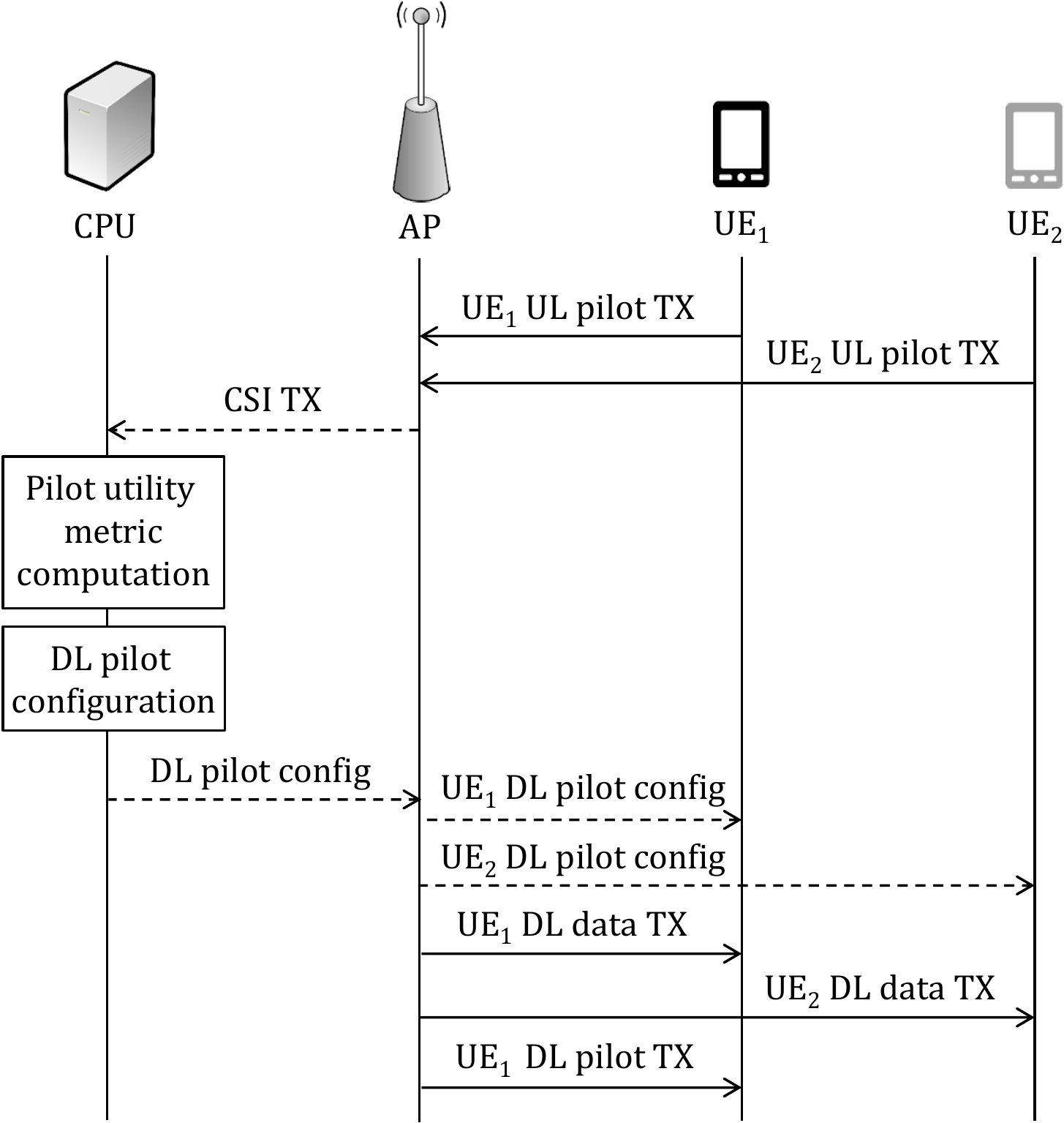}
    \caption{In the utility-based DL pilot assignment strategy, the CPU collects from the APs all the information needed to compute the pilot utility metric. Afterwards, it sends the DL pilot config messagge to the APs to inform which UE is assigned an orthogonal DL pilot.}\label{fig:diagram}
\end{figure}

\begin{figure}[!t]
    \centering
    \includegraphics[width=.95\linewidth]{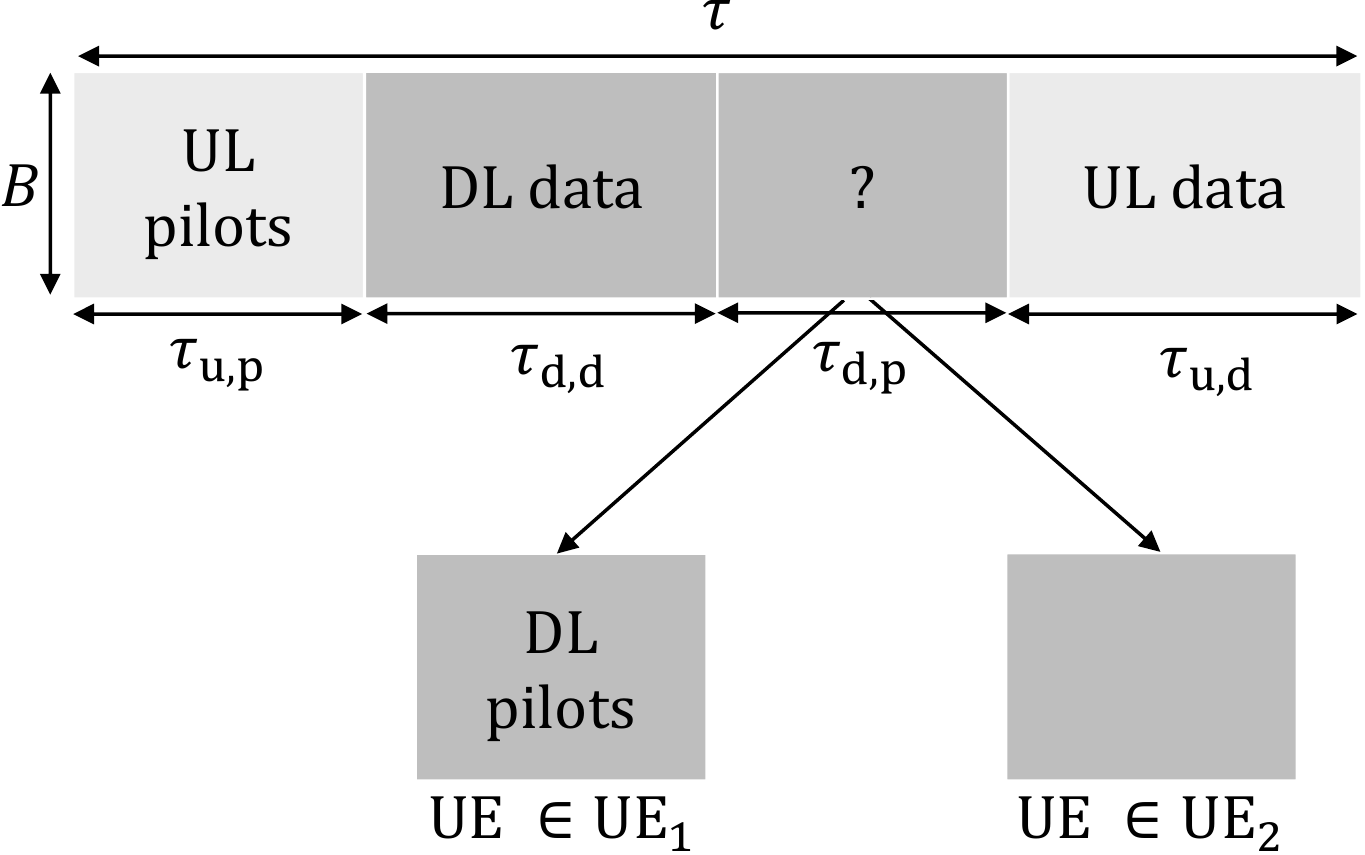}
    \caption{In the utility-based DL pilot assignment strategy, the DL training phase is performed only for those users that need to estimate the instantaneous CSI the most (UE$_1$, in this example). The rest of the UEs (UE$_2$, in this example) rely on statistical CSI. In any case, $\tau_{\textrm{d,p}}$ symbols cannot be used for data, in order not to create data-to-pilot interference.}\label{fig:TDDframePS}
\end{figure}

The detailed description of the proposed scheme is illustrated from the signaling diagram in~\Figref{fig:diagram}. All the UEs send their own UL pilot at the beginning of the coherence interval, as shown in~\Figref{fig:dlpilots}. The AP estimates the UL channel and sends the corresponding CSI to the CPU. The CPU collects this information from each AP, and computes the pilot utility metric for each user. Based on this information, the CPU groups the UEs to two categories: $(i)$ UEs requiring DL pilots, having a pilot utility metric above a predetermined threshold (UE$_1$ in~\Figref{fig:diagram}); $(ii)$ UEs not requiring DL pilots, having a pilot utility metric below a predetermined threshold (UE$_2$ in~\Figref{fig:diagram}). Once these two groups are defined, the CPU sends to all the APs the \textit{DL pilot configuration} message, which conveys the DL-pilot-to-UE mapping. Hence, the AP forwards part of this message (it might be just one bit) to each UE. After the data transmission, the APs send the orthogonal DL pilots only to the UEs of the first category (UE$_1$). 
UE$_1$ knows a priori the switching time from the DL data to the DL training phase since the TDD frame structure is determined on large time scale. If a DL pilot is assigned, the UE estimates the DL channel based on the DL pilot and demodulates the data on the remaining resources assigned to the DL transmission. If a DL pilot is not assigned, the UE assumes that the DL channel can be approximated by a constant, and it demodulates the data on all resources assigned to the DL transmission by exploiting the statistical CSI knowledge. No resources are assigned to UE$_2$ during the DL training phase, in order to avoid data-to-pilot interference. The TDD frame for the proposed scheme is depicted in~\Figref{fig:TDDframePS}.

The achievable DL rate, subject to the assumption that the UEs transmit Gaussian message-bearing symbols, for the utility-based DL pilot assignment scheme is given by
\begin{align}
R_k^{\textrm{ubPA}} = 
\begin{cases}
R_k^{\textrm{iCSI}}, \quad \text{if UE~} k \in \text{UE}_1,   \\
R_k^{\textrm{sCSI}}, \quad \text{otherwise}.
\end{cases}
\end{align}
\subsection{Pilot utility metric}

The pilot utility metric can be defined in different manner according to the purposes. It is a function of the UE mobility, the channel hardening degree, and the CSI. In general, it must guarantee an orthogonal DL pilot to: $(i)$ high-speed UEs, which necessarily need the instantaneous CSI to face the CSI aging; $(ii)$ low/moderate-speed UEs experiencing a low degree of channel hardening, which require instantaneous CSI to face the unreliability of the statistical CSI. On the other hand, all low/moderate-speed UEs for which the channel hardens sufficiently will not require DL training, and they may rely only on statistical CSI. Hence, UEs may be assigned with an orthogonal DL pilot if the Doppler spread value is above a pre-determined threshold and/or the channel hardening degree is below a pre-determined threshold. Such a pilot utility metric can be defined as
\begin{equation}
\textrm{pu}_k = wD_k + (1-w)(1-\textrm{ChD}_k),
\end{equation}
where $D_k \in [0,1]$ is the Doppler spread value, $\textrm{ChD}_k$ indicates the \textit{channel hardening degree} at the $k$th UE, and $w \in [0,1]$ is the weight to prioritize differently the UE mobility and the channel hardening degree. 

The channel hardening degree is defined as the ratio between the instantaneous channel gain and its average value. By following a similar methodology as in~\cite{ngo2017nodownlinkpilots}, we obtain the channel hardening degree expression at the $k$th UE for a cell-free massive MIMO system,
\begin{equation}\label{eq:ChD}
\textrm{ChD}_k=1-\frac{\text{Var}\left\{\sum\nolimits^M_{m=1}|g_{mk}|^2\right\}}{\left(\mathbb{E}\left\{\sum\nolimits^M_{m=1} |g_{mk}|^2\right\}\right)^2}.
\end{equation}
The larger $\textrm{ChD}_k$ is, the more the channel hardens for UE $k$.
 
The pilot utility metric may also account for the UE priority in terms of \textit{Quality of Service} (QoS) requirements. This needs the APs collect and send to the CPU the QoS requirements for each UE, and the CPU determines the UE priorities accordingly. In this case, the pilot utility metric is
\begin{align} \label{eq:ChDpu}
\textrm{pu}_k = 
\begin{cases}
wD_k + (1-w)(1-\textrm{ChD}_k)+\alpha_k,\quad\text{or} \\[10pt]
\alpha_k[wD_k + (1-w)(1-\textrm{ChD}_k)],
\end{cases}
\end{align}
where $\alpha_k \in [0,1]$ is the UE priority. The larger $\alpha_k$ is, the more QoS the UE $k$ requires.
 
Alternatively to the channel hardening degree estimation, a DL pilot may be assigned to those users that would increase their rates the most by taking advantage of the DL pilot. In this case, the pilot utility metric may be defined as a function of the achievable rates (or throughputs) estimated by using the CSI knowledge at each AP. The following pilot utility metrics
\begin{align} \label{eq:puk}
\textrm{pu}_k =
\begin{cases}
\alpha_k\left[wD_k + (1-w)\left(R_k^{\textrm{iCSI}} - R_k^{\textrm{sCSI}}\right)\right], \\[10pt]
\alpha_k\left[wD_k + (1-w)\left(T_k^{\textrm{iCSI}} - T_k^{\textrm{sCSI}}\right)\right], \\[10pt]
\alpha_k\left[wD_k + (1-w)\left(\frac{R_k^{\textrm{iCSI}} - R_k^{\textrm{sCSI}}}{R_k^{\textrm{iCSI}}}\right)\right], \\[10pt]
\alpha_k\left[wD_k + (1-w)\left(\frac{T_k^{\textrm{iCSI}} - T_k^{\textrm{sCSI}}}{T_k^{\textrm{iCSI}}}\right)\right], \\[10pt]
\alpha_k\left[wD_k + (1-w)\left(R_k^{\textrm{sCSI}}\right)\inv \right],
\end{cases}
\end{align} 
are valid alternatives, according to different aims: absolute rate increase, absolute throughput increase, relative rate increase, relative throughput increase, and low-rate prioritization, respectively.   
The \textit{per-user net throughput} (bit/s) is proportional to the per-user rate, and takes into account the performance loss due to the DL and UL pilots transmission (as they subtract resources to the data). For a symmetric TDD frame, i.e., $\tau_{\textrm{u,d}} = \tau_{\textrm{d,d}}$, it is given by
\begin{equation} \label{eq:throughput}
T_k = \frac{B}{2} \left(\frac{1-\tau_{\textrm{p}}}{\tau}\right) R_k,
\end{equation}
where $B$ is the bandwidth, and $\tau_{\textrm{p}}$ is the pilot overhead.
\section{Numerical Results} \label{sec:NumericalResults}

Our simulations aim to compare the performance, in terms of DL net sum throughput and average DL net per-user throughput, provided by our solution and the existing solutions of the state-of-the-art. 
The DL net sum throughput, measured in bits/sec, is equal to the sum of the DL net throughputs per UE defined in~\eqref{eq:throughput}, and it is given by $T = \sum\nolimits_{k=1}^K T_k$. Hence, the average DL net per-user throughput is $T_{\textrm{avg}} = T/K$.
\subsection{Simulation Scenario}
Assuming the channel model in~\eqref{eq:channelmodel}, we describe the path-loss and shadow-fading considered in our performance evaluation. The coefficients $\{\beta_{mk}\}$ are defined as follows
\begin{equation}
\label{eq:beta}
\beta_{mk} = \text{PL}_{mk} \cdot 10^{\frac{\sigma_{sh}z_{mk}}{10}},
\end{equation}  
where $\text{PL}_{mk}$ represents the path loss, $10^{\frac{\sigma_{sh}z_{mk}}{10}}$ is the shadow-fading with standard deviation $\sigma_{sh}$, and $z_{mk}\sim\mathcal{N}(0,1)$, i.e., we assume uncorrelated shadow-fading. 
The path-loss is modeled as three-slope~\cite{ngo2017cellfree}, where the loss exponent can assume three different values according to the UE-to-AP distance. More specifically, the loss exponent is~$3.5$~if the distance between UE $k$ and AP $m$ ($d_{mk}$) is greater than a first reference distance~$d_1$; it is equal to~$2$~if $d_{mk}$ is greater than a second reference distance~$d_0$ and less or equal to $d_1$. In the case $d_{mk}$ is less than~$d_0$~the loss exponent is $0$. The chosen propagation model is the Hata-COST231. Hence, the path-loss in dB is given by
\begin{equation}
\label{eq:3slopePL}
\text{PL}_{mk}\!=\!
\begin{cases}
\!-L\!-35\!\log_{10}\!(d_{mk})\!&\!\text{if }\!d_{mk}\!>\!d_1, \\
\!-L\!-15\!\log_{10}\!(d_1)\!-\!20\!\log_{10}\!(d_{mk})\!&\!\text{if }\!d_0\!<\!d_{mk}\!\leq\!d_1, \\
\!-L\!-15\!\log_{10}\!(d_1)\!-\!20\!\log_{10}\!(d_0)\!&\!\text{if }\!d_{mk}\!\leq\!d_0, \\
\end{cases}
\end{equation}
where 
\begin{align}
\label{eq:hatacost}
L &= 46.3\!+\!33.9\log_{10}(f)\!-\!(1.1\log_{10}(f)-0.7)h_{\text{UE}}\! \nonumber \\
&\quad -\!13.82\log_{10}(h_{\text{AP}})\!+\!(1.56\log_{10}(f)\!-\!0.8).
\end{align}
In our simulations, $d_1=50$ meters, $d_0=10$ meters, the carrier frequency $f$ (in MHz) is $2$ GHz, the height of the UE and AP antennas are~$1.65$~meters and~$5$~meters, respectively. In addition, the shadow-fading coefficients have standard deviation 8 dB. 

The other simulation parameters are set as follows: $M=200$~APs and $K=50$~UEs are randomly and uniformly distributed within a square of size $1 \text{ km}^2$; the bandwidth $B$ is $20$ MHz; the antenna gains are $0$ dBi; the noise figure (both for UL and DL) is 9 dB; the radiated power for data and pilot is $200$ mW for APs, and $100$ mW for UEs ($\bar{\rho}_{\textrm{d}}$, $\bar{\rho}_{\textrm{d,p}}$, $\bar{\rho}_{\textrm{u}}$, $\bar{\rho}_{\textrm{u,p}}$, respectively). The corresponding normalized transmit SNRs ($\rho_{\textrm{d,p}}$, $\rho_{\textrm{d}}$, $\rho_{\textrm{u}}$, $\rho_{\textrm{u,p}}$), defined in Section~\ref{sec:sysmodel}, are obtained by dividing the radiated powers by the noise power, which is given by
\begin{align*}
\text{noise power} = B \times k_B \times T_0 \times \text{noise figure} \text{ (W)},
\end{align*}
where $k_B$ is the Boltzmann constant, and $T_0 = 290$ (Kelvin) is the noise temperature.
We take the length of the TDD frame $\tau = 200$ symbols, which corresponds to a coherence bandwidth of 200 kHz and a
time-slot of 1 ms. To simulate a cell-free network topology (i.e., no cell-edge effects), we wrap the simulation area around with eight twin neighbor areas. The UL and DL orthogonal pilot sequences are randomly assigned to the users. Finally, the power control coefficients $\{\eta_{mk}\}$ are set by performing \textit{max-min fairness power control}, as described in~\cite[Section IV-B]{ngo2017cellfree}. The simulation settings are summarized in Table~\ref{tab:settings}.

\begin{table}[!t]
\caption{Simulation settings}
\label{tab:settings}
\centering
\renewcommand{\arraystretch}{1.2}
\begin{tabular}{l | l || l | l}
Description & Value & Description & Value \\
\hline
APs/UEs & unif. rand. distr. & $M$ & $200$ \\
simulation area & 1 km$^2$ & $K$ & $50$ \\
$\tau$ & 200 symbols & time-slot & $1$ ms\\ 
path-loss & three-slope & $d_1, d_0$ & $50, 10$ m \\ 
carrier frequency & 2 GHz & bandwidth $B$ & 20 MHz \\
shadow-fading & uncorrelated & antenna gain & 0 dBi \\
shadow-fading std & 8 dB & noise figure & 9 dB \\
AP antenna height & 5 m & $\bar{\rho}_{\textrm{d}}$, $\bar{\rho}_{\textrm{d,p}}$ & 200 mW \\
UE antenna height & 1.65 m & $\bar{\rho}_{\textrm{u}}$, $\bar{\rho}_{\textrm{u,p}}$ & 100 mW
\end{tabular}
\end{table}

\subsection{Performance Evaluation}

\begin{figure}[!t]
\centering
\includegraphics[width=\linewidth]{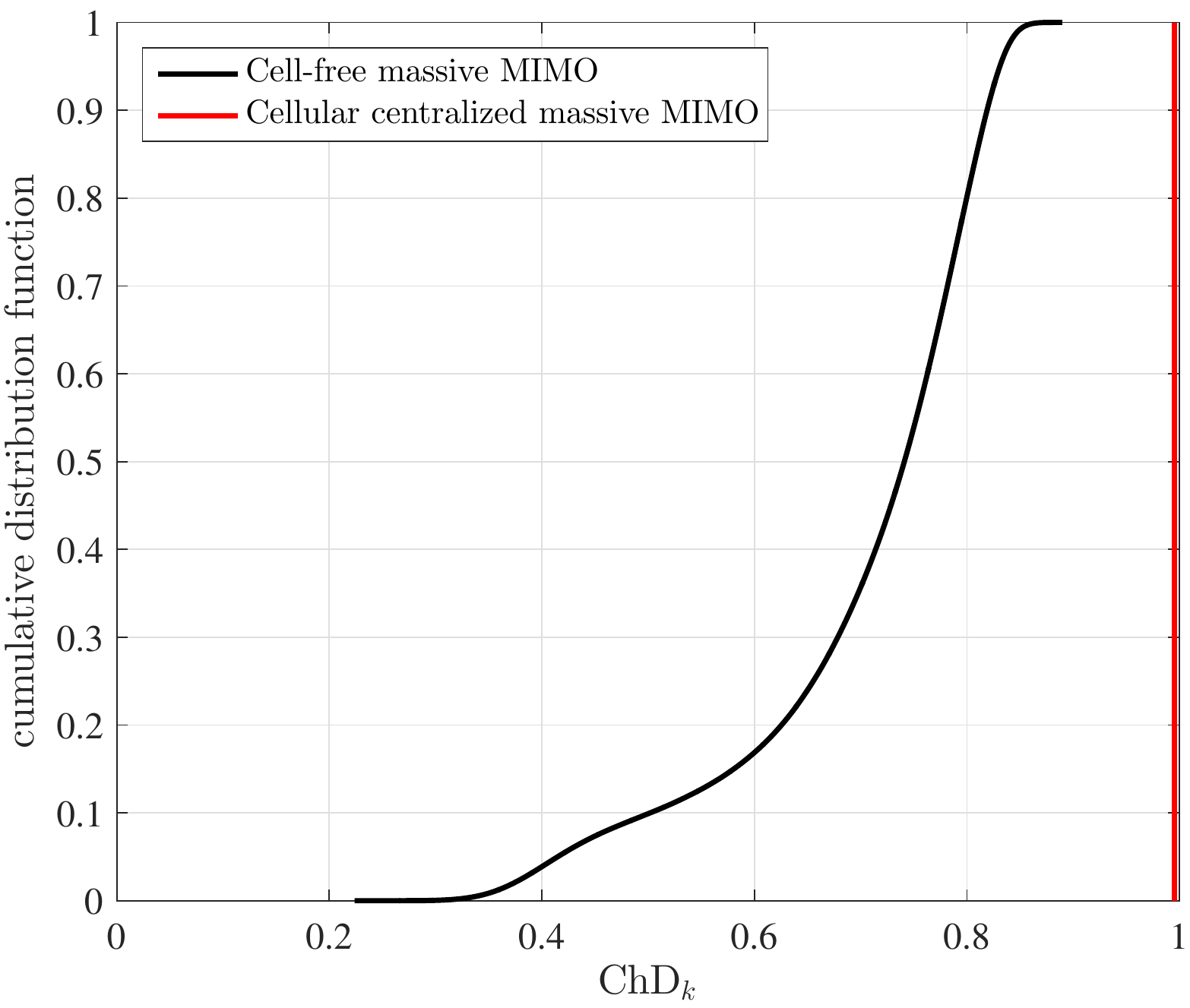}
\caption{The $\textrm{ChD}$ in cell-free massive MIMO is much lower compared to that in conventional massive MIMO. Here, we assume 200 APs randomly distributed in an area of 1 km$^2$, and i.i.d. Rayleigh fading channel.}
\label{fig:cdfChD}
\end{figure}

Firstly, we measure the channel hardening degree in such a cell-free massive MIMO system. 
The cumulative distribution function (cdf) of the channel hardening degree is shown in~\Figref{fig:cdfChD}, for cell-free massive MIMO and the cellular centralized massive MIMO. The cdf gives more insights about the distribution of the channel hardening degree given the random variables of the AP/UE positions in the deployment area. In cell-free massive MIMO the channel hardening degree is given by~\eqref{eq:ChD}. Instead, in conventional massive MIMO, for i.i.d. Rayleigh fading channels, the channel hardening degree simply scales as $1/M$~\cite{ngo2017nodownlinkpilots}, and in this case it is given by $1-1/M = 0.995$, regardless of the AP/UE position. \Figref{fig:cdfChD} confirms that in cell-free massive MIMO the channel hardening degree is significantly lower than in conventional massive MIMO.  

Next, we introduce an example to show the benefits provided by our solution with respect to the state-of-the-art. Let us consider the following case studies:
\begin{enumerate}
\item \textit{statistical CSI} (\textrm{sCSI}), prior art: the coherence interval is structured as in~\Figref{fig:ulpilots}. No DL training is performed and all the active UEs rely only on statistical CSI to decode data. Each UE is assigned with an orthogonal UL pilot. Hence, $\tau_\textrm{u,p} = 50$, $\tau_\textrm{d,p} = 0 \implies \tau_\textrm{p} = 50$.
\item \textit{instantaneous CSI} (\textrm{iCSI}), prior art: the coherence interval is structured as in~\Figref{fig:dlpilots}. The DL training is performed for all UEs. Each UE is assigned with an orthogonal UL pilot and receives an orthogonal DL pilot from which it estimates the instantaneous CSI. Hence, $\tau_\textrm{u,p} = \tau_\textrm{d,p} = 50 \implies \tau_\textrm{p} = 100$.
\item \textit{utility-based pilot assignment} (\textrm{ubPA}), proposed scheme: the coherence interval is structured as in~\Figref{fig:TDDframePS}. Each UE is assigned with an orthogonal UL pilot. All the UEs having a pilot utility metric exceeding a predetermined threshold are assigned with an orthogonal DL pilot. The remaining UEs rely only on statistical CSI.
\begin{remark}
In a realistic scenario, the assignment of UL/DL resources occurs in a quasi-static fashion, since a dynamic re-allocation of pilots, frame by frame, requires a huge amount of control signaling to let the UEs know the exact UL/DL switching time. This also  introduces delay. Therefore, the predetermined threshold is actually a design setting varying in large time scale. 
\end{remark}
In view of the above, let us assume the following UL/DL resources assignment: $\tau_\textrm{u,p}\! =\! 50, \tau_\textrm{d,p} \!=\! 25\! \implies\! \tau_\textrm{p}\!=\!75$. 
Hence, according to the proposed scheme, only the $25$ UEs out of $50$ with the highest pilot utility metric receive an orthogonal DL pilot. The remaining $25$ UEs have no allocated resource during the DL training phase, in order to avoid further interference, as shown in~\Figref{fig:TDDframePS}.
\end{enumerate}
With the purpose to maximize the DL net sum throughput, we set the pilot utility metric as in the first expression of~\eqref{eq:puk} (absolute rate increase). For simplicity, we finally assume that the length of the TDD frame supports the speed of all the active users in the network, and all the UEs have equal QoS priority, that is $w=0, \alpha_k = 1$, respectively.
\begin{remark}
Since the number of UEs that can be served is proportional to the time spent sending pilots, while the sum throughput is proportional to the number of UEs served, it follows that a good rule to allocate the UL/DL pilot resources is simply setting the number of symbols spent on transmission of UL and DL pilots not exceeding half the length of the coherence interval. 
\end{remark}

According to this, and the simulation scenario, i.e., $\tau=200$, the maximum number of symbols dedicated for the pilots is $\tau_{\textrm{p}} = 100$. Under this assumption, in the \textrm{iCSI} case all the UEs are served in the current frame. Conversely, if the number of UEs would be $K > 50$, then $\tau_{\textrm{p}} > 100$, and in the \textrm{iCSI} case the exceeding UEs would be scheduled in the next frame.     
\begin{figure}[!t]
\centering
\includegraphics[width=\linewidth]{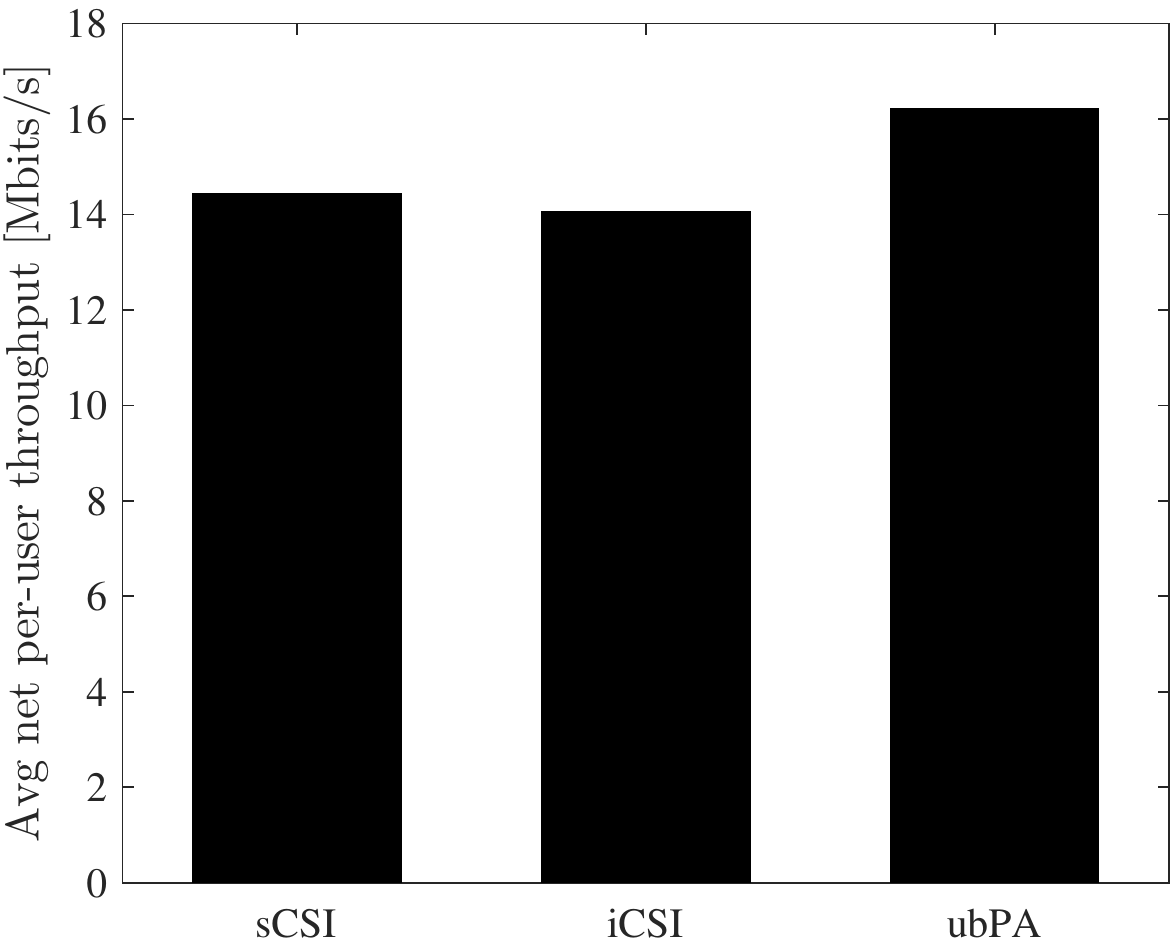}
\caption{Performance comparison between the \textrm{sCSI}, \textrm{iCSI}, and \textrm{ubPA} scheme. The utility-based pilot assignment strategy guarantees higher average DL net per-user throughput than the state-of-the-art schemes.}
\label{fig:barsT}
\end{figure}

The simulation results in~\Figref{fig:barsT} show that \textrm{ubPA} scheme performs better than the prior art schemes in terms of average net per-user throughput, providing about 12\% improvement over case \textrm{sCSI}, and about 16\% improvement over case \textrm{iCSI}.

\begin{figure}[!t]
\centering
\includegraphics[width=\linewidth]{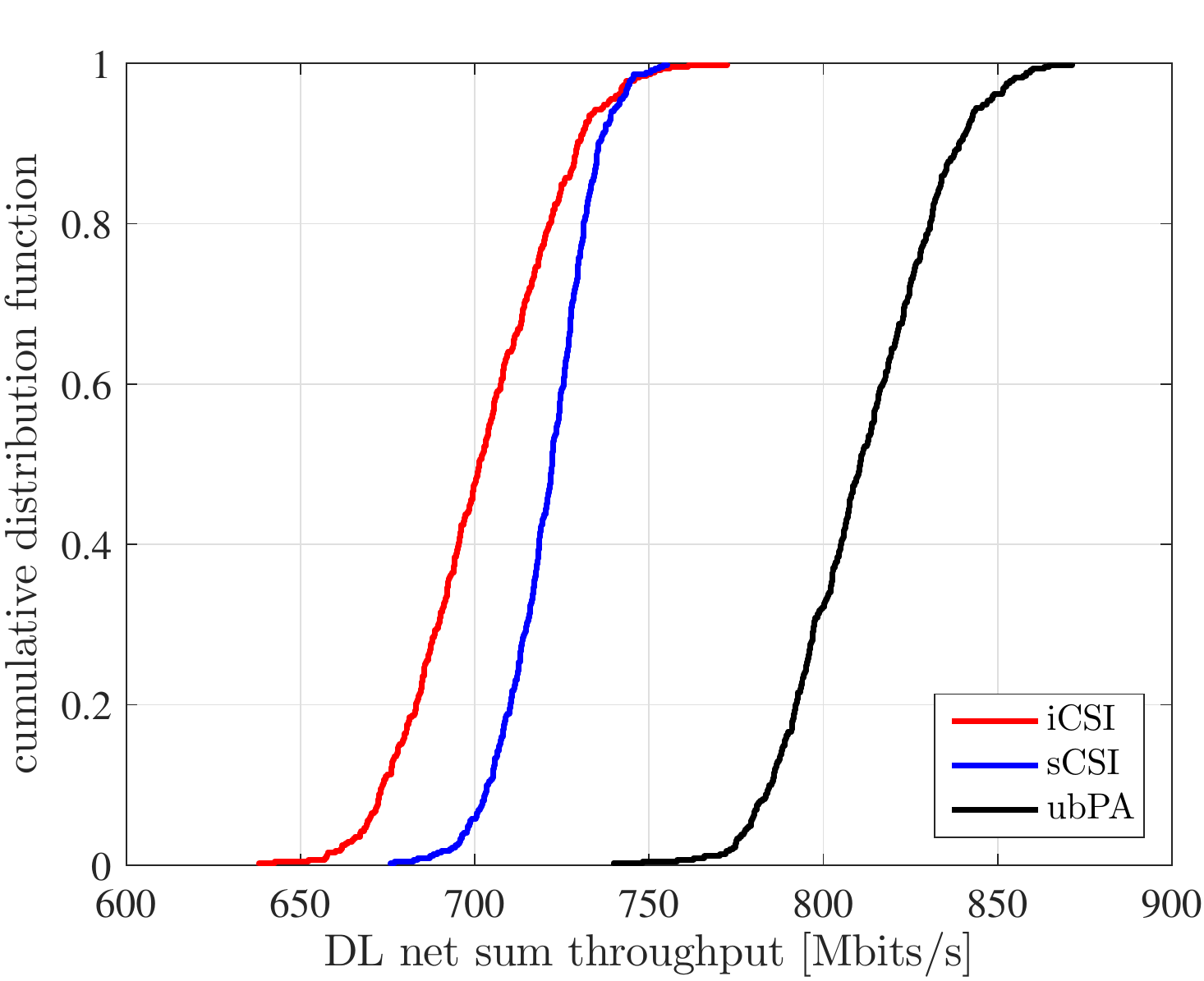}
\caption{Cdf of the DL net sum throughput provided by the \textrm{sCSI}, \textrm{iCSI}, and \textrm{ubPA} scheme. The utility-based pilot assignment strategy guarantees higher DL net sum throughput than the state-of-the-art schemes.}
\label{fig:cdfT}
\end{figure}

The gain introduced by the proposed scheme in terms of DL net sum throughput is shown in~\Figref{fig:cdfT}. As we can see, the \textrm{ubPA} scheme provides a gain over scheme \textrm{sCSI} by about 11\%, 12\% and 13\% at the 5th percentile, median and 90th percentile, respectively. These gains are larger over scheme \textrm{iCSI}: 17\% at the 5th percentile, 16\% at the median point and 15\% at the 90th percentile. The results also show that in such a scenario with many APs and UEs, and short coherence interval, transmitting one orthogonal DL pilot per UE (scheme \textrm{iCSI}) is less efficient than letting the UEs rely on statistical CSI (scheme \textrm{sCSI}). In fact, the performance loss due to the pilots transmission overhead overcomes the benefits, in terms of higher effective SINRs, introduced by the DL training.

Lastly, we compare the DL net sum throughputs provided by three different pilot utility metrics: $(i)$ absolute rate increase, first expression in~\eqref{eq:puk}; $(ii)$ absolute throughput increase, second expression in~\eqref{eq:puk}; $(iii)$ $\textrm{ChD}$-aware pilot utility, second expression in~\eqref{eq:ChDpu}.  
\begin{figure}[!t]
\centering
\includegraphics[width=\linewidth]{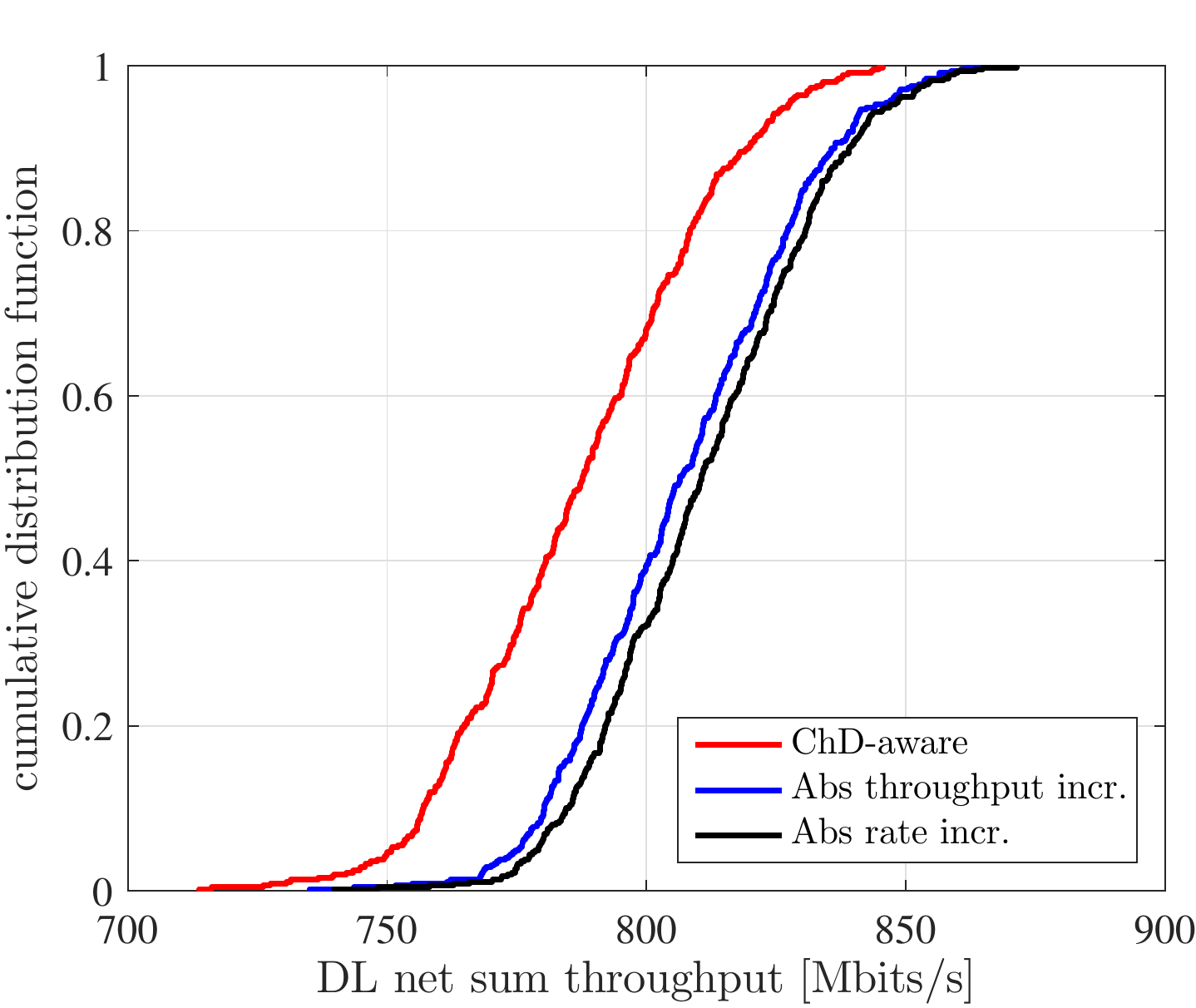}
\caption{Cdf of the DL net sum throughput provided by the absolute rate and throughput increase, and $\textrm{ChD}$-aware pilot utility metric. The absolute rate increase approach guarantees higher performance than other approaches.}
\label{fig:cdfPU}
\end{figure}
As shown in~\Figref{fig:cdfPU}, the absolute rate increase pilot utility metric guarantees the highest DL net sum throughput even though the gain over the absolute throughput increase approach is negligible. 

\balance

On the other hand, the gain with respect to the $\textrm{ChD}$-aware approach may reach up to 5\%. Estimating the rates/throughputs can give more information about the UE's real need of a DL pilot rather than estimating the $\textrm{ChD}$. In fact, the pilot utility metric based on the rate/throughput estimates takes into account both the $\textrm{ChD}$ by estimating the beamforming uncertainty gain, and the amount of inter-user interference. Both factors affect user's performance and they need to be jointly evaluated when assigning the DL pilots.     

\section{Conclusion}
In this paper, we proposed a strategy for orthogonal DL pilot assignment based on the so-called pilot utility metric, which is a function of the channel hardening degree or the rate estimate, user mobility, user QoS priority, and CSI.  

The proposed strategy consists in assigning the orthogonal DL pilots only to the users that benefits the most when decoding data by exploiting the instantaneous CSI. In general, this approach guarantees an orthogonal DL pilot to users with high mobility, to cope with the CSI aging, and users with low/moderate mobility experiencing low channel hardening, for which the reliance on statistical CSI in the decoding yields poor performance. The utility-based pilot assignment scheme guarantees higher DL per-user and sum throughput, better support for high-speed users and shorter coherent interval compared to the state-of-the-art.

\bibliographystyle{IEEEtran}
\bibliography{IEEEabrv,mybib}
\end{document}